%% file: ms.tex
\documentclass[12pt,preprint]{aastex}

\usepackage{graphicx}
\usepackage{natbib}

\newcommand{\La}{\mbox{${\rm Ly\alpha}$}}
\newcommand{\Nh}{\mbox{$N_{\rm H}$}}
\newcommand{\Rwd}{\mbox{$R_{\rm wd}$}}
\newcommand{\Mwd}{\mbox{$M_{\rm wd}$}}
\newcommand{\Twd}{\mbox{$T_{\rm wd}$}}
\newcommand{\Tspot}{\mbox{$T_{\rm spot}$}}

\newcommand{\Msun}{\mbox{$M_{\odot}$}}
\newcommand{\FLP}{\mbox{$\mathrm{F28\times50LP}$}}
\def\FLUXARCSEC '\:ergs\:cm sup -2 s sup -1 arcsec sup -2'
\def\COUNTS '\:s sup -1 '
\def\CPM '\:s sup -1 arcmin sup -2'

\def\BT{B_T}

\def\FLUXARCSEC{\:{\rm ergs\:cm^{-2}\:s^{-1}\:arcsec^{-2}}}
\def\COUNTS{\:{\rm cts\:s^{-1}}}
\def\CPM{\:{\rm s^{-1}\:arcmin^{-2}}}

\def\HEiiL{He\thinspace {\footnotesize II} $\lambda1640$}

\def\NvL{N\thinspace {\footnotesize V} $\lambda\lambda1239,1243$}

\def\CiiLxiii{C\thinspace {\footnotesize II} $\lambda1333$}

\def\CiiiLxi{C\thinspace {\footnotesize III} $\lambda1176$}
\def\CivL{C\thinspace {\footnotesize IV} $\lambda\lambda1548,1551$}

\def\SIiiiL{Si\thinspace {\footnotesize III} $\lambda\lambda1295$--1303}
\def\SIivL{Si\thinspace {\footnotesize IV} $\lambda\lambda1394,1403$}

\def\SIiiiLa{Si\thinspace {\footnotesize III} $\lambda1207$}
\def\SIivLa{Si\thinspace {\footnotesize IV} $\lambda1394$}
\def\SIivLb{Si\thinspace {\footnotesize IV} $\lambda1403$}
\def\CivLL{C\thinspace {\footnotesize IV} $\lambda\lambda1548,51$}
\def\SIiiiLL{Si\thinspace {\footnotesize III} $\lambda\lambda1295,03$}
\def\NvLL{N\thinspace {\footnotesize V} $\lambda\lambda1239,43$}
\def \HEiil{He\thinspace {\footnotesize II}}
\def \Civl{C\thinspace {\footnotesize IV}}
\def\SIivl{Si\thinspace {\footnotesize IV}}
\def\Nvl{N\thinspace {\footnotesize V}}

\begin{document}


\title{Far-ultraviolet Spectroscopy of Magnetic Cataclysmic Variables \altaffilmark{1}}

\author{Sofia Araujo-Betancor}
\affil{Space Telescope Science Institute, 3700 San Martin Drive,
Baltimore, MD\,21218, USA}
\email{araujo@stsci.edu}
\author{Boris T. G\"ansicke} 
\affil{Department of physics, University of Warwick, Coventry,
  CV4\,7AL, UK}
\email{Boris.Gaensicke@warwick.ac.uk}
\author{Knox S. Long}
\affil{Space Telescope Science Institute, 3700 San Martin Drive,
Baltimore, MD\,21218, USA}
\email{long@stsci.edu}
\author{Klaus Beuermann} \affil{Universit\"ats-Sternwarte,
Geismarlandstrasse 11, G\"ottingen 37083, Germany} 
\email{beuermann@uni-sw.gwdg.de}
\author{Domitilla de Martino} \affil{Osservatorio di Capodimonte, Via
Moiariello 16, Naples I-80131, Italy}
\email{demartin@na.astro.it}
\author{Edward M. Sion} \affil{Department of Astronomy and
Astrophysics, Villanova University, 800 Lancaster Avenue, Villanova,
PA\,19085, USA}
\email{edward.sion@villanova.edu}
\and
\author{Paula Szkody} \affil{Department of Astronomy, University of
Washington, Seattle, WA\,98195, USA}
\email{szkody@astro.washington.edu}


\altaffiltext{1}{Based on observations made with the NASA/ESA Hubble Space
Telescope, obtained at the Space Telescope Science Institute, which is
operated by the Association of Universities for Research in Astronomy,
Inc., under NASA contract NAS 5-26555.}

\begin{abstract}
We have obtained \textit {HST}/STIS data for a total of eleven polars
as part of a program aimed to compile a homogeneous database of
high-quality far-ultraviolet (FUV) spectra for a large number of
cataclysmic variables (CVs). Of the eleven polars, eight were found in
a state of low accretion activity (V347\,Pav, VV\,Pup, V834\,Cen,
BL\,Hyi, MR\,Ser, ST\,LMi, RX\,J1554.2+2721 and V895\,Cen) and three
in a state of high activity (CD\,Ind, AN\,UMa and UW\,Pic).  The STIS
spectra of the low-state polars unambiguously reveal the photospheric
emission of their white dwarf (WD) primaries. We have used pure
hydrogen WD models to fit the FUV spectra of the low-state systems
(except RX\,J1554.2+2721, which is a high-field polar) in order to
measure the WD effective temperatures. In all cases, the fits could be
improved by adding a second component, which is presumably due to
residual accretion onto the magnetic pole of the WD. The WD
temperatures obtained range from 10\,800\,K to 14\,200\,K for $\log g
= 8.0$. Our analysis more than doubles the number of polars with
accurate WD effective temperatures. Comparing the WD temperatures of
polars to those of non-magnetic CVs, we find that at any given orbital
period the WDs in polars are colder than those in non-magnetic CVs.
The temperatures of polars below the period gap are consistent with
gravitational radiation as the only active angular momentum loss
mechanism. The differences in WD effective temperatures between polars
and non-magnetic CVs are significantly larger above the period gap,
suggesting that magnetic braking in polars might be reduced by the
strong field of the primary. We derive distance estimates to the
low-state systems from the flux scaling factors of our WD model
fits. Combining these distance measurements with those from the
literature, we establish a lower limit on the space density of polars
of $1.3\times10^{-6}\,\rm pc^{-3}$.

\end{abstract}

\section{Introduction}
Polars, also called AM\,Her stars, are cataclysmic variables (CVs)
containing a white dwarf (WD) with a strong magnetic field ($ \rm B
\ge 10\,MG$) and a Roche lobe filling late type star (the secondary)
that is losing mass to the WD through the inner Lagrangian point
($L_1$). For the typical $B \sim 10 - 80$\,MG fields in polars, the
material travels on a ballistic trajectory from $L_1$ to a point where
the magnetic pressure exceeds the ram pressure of the accretion
stream. At that point, the material is captured by the field and falls
along the field lines to the magnetic pole or poles of the WD. A disk
does not form because the material is captured before it reaches the
stream circularization radius. The strong magnetic interaction between
the WD and the donor star also results in the synchronization of the
WD with the orbital period of the system. Due to the absence of a
disk, polars do not show the outbursts characteristic of dwarf
novae. Instead, they undergo deep low states on irregular timescales
of months to years, up to three magnitudes fainter than their bright
(high) states, during which the WD accretes at an extremely low
rate. Since there is no disk to buffer the material before it reaches
the WD, variations in accretion rate onto the WD directly reflect
changes in the mass loss rate of the secondary \citep[possibly due to
star spots covering $L_1$;][]{hessmanetal00-1}. Even though mass
transfer via Roche lobe overflow may cease during a low state, the WD
might still accrete some material from the donor star wind, as occurs
in post common envelope binaries \citep[see
e.g.\,][]{odonoghueetal03-1}. The sources of ultraviolet (UV) emission
that dominate in the high state - a hot polar cap near the footpoint
of the accretion column, contributing most in the far-UV (FUV), and
the X-ray illuminated accretion stream - greatly weaken or completely
disappear during low-states. In this situation, the WD photosphere is
the main source of FUV-radiation \citep[e.g][]{gaensickeetal95-1,
stockmanetal94-1, gaensicke98-2, gaensickeetal00-1, rosenetal01-1,
schwopeetal02-1} and the effective temperature of the WD can be
directly measured.

With typical effective temperatures in the range $10\,000-50\,000$\,K,
the best-suited wavelength range to observe CV\,WDs is in the region
from $1500-3000$\AA. As a consequence, much of the efforts on studying
CV\,WDs were directed to the UV regime.  The first measurements of WD
effective temperatures in CVs were carried out with the
\textit{International Ultraviolet Explorer} (\textit{IUE}) two decades
ago \citep{mateo+szkody84-1, panek+holm84-1}. As a result of the
\textit{IUE} observations, \citet{sion85-1, sion91-1} recognized that
WDs in CVs were hotter than field WDs of comparable age, and suggested
that compressional heating by the accreted material could account for
this fact. More recently, \citet{townsley+bildsten02-2,
townsley+bildsten03-1, townsley+bildsten04-1} have carried out
detailed studies of the long term effects of accretion on the thermal
structure of the core and the envelope, including low-level nuclear
burning, and have determined a relationship between the long-term
average accretion rate and the photospheric temperature of the WD.
This makes the measurement of CV\,WD temperatures an extremely
important issue, since the secular mean of the mass accretion rates is
key to understanding the evolutionary history of CVs.

A total of 23 polars were observed with \textit{IUE} during its long
and successful history, and these observations allowed a number of
important insights into the accretion process of magnetic CVs
\citep[see e.g.\,][for a review]{demartino98-1}. Only a handful of
these systems were bright enough to obtain estimates of their WD
temperatures \citep[e.g.\,][]{heise+verbunt88-1, gaensickeetal95-1,
gaensickeetal00-1}, the vast majority of polars were too faint for
\textit{IUE} to obtain UV spectroscopy of their WDs during low
states. The larger aperture of the \textit{Hubble Space Telescope}
(\textit{HST}) allowed observations of fainter polars using the Faint
Object Spectrograph (FOS) and the Goddard High Resolution Spectrograph
(GHRS), but the number of different polars (and CVs in general)
measured remained quite small \citep{stockmanetal94-1,
demartinoetal98-1, rosenetal01-1, schwopeetal02-1}. To address this
problem, we undertook an \textit{HST} snapshot survey with the goal of
obtaining reasonable high quality FUV spectra of a substantial sample
of CVs of all subclasses \citep{gaensickeetal03-1}. Here we report on
the observations of the eleven polars obtained as part of this survey.

\section{Observations and Data Reduction}
\label{s-observations}

\begin{figure*}
\centerline{\includegraphics[width=18cm]{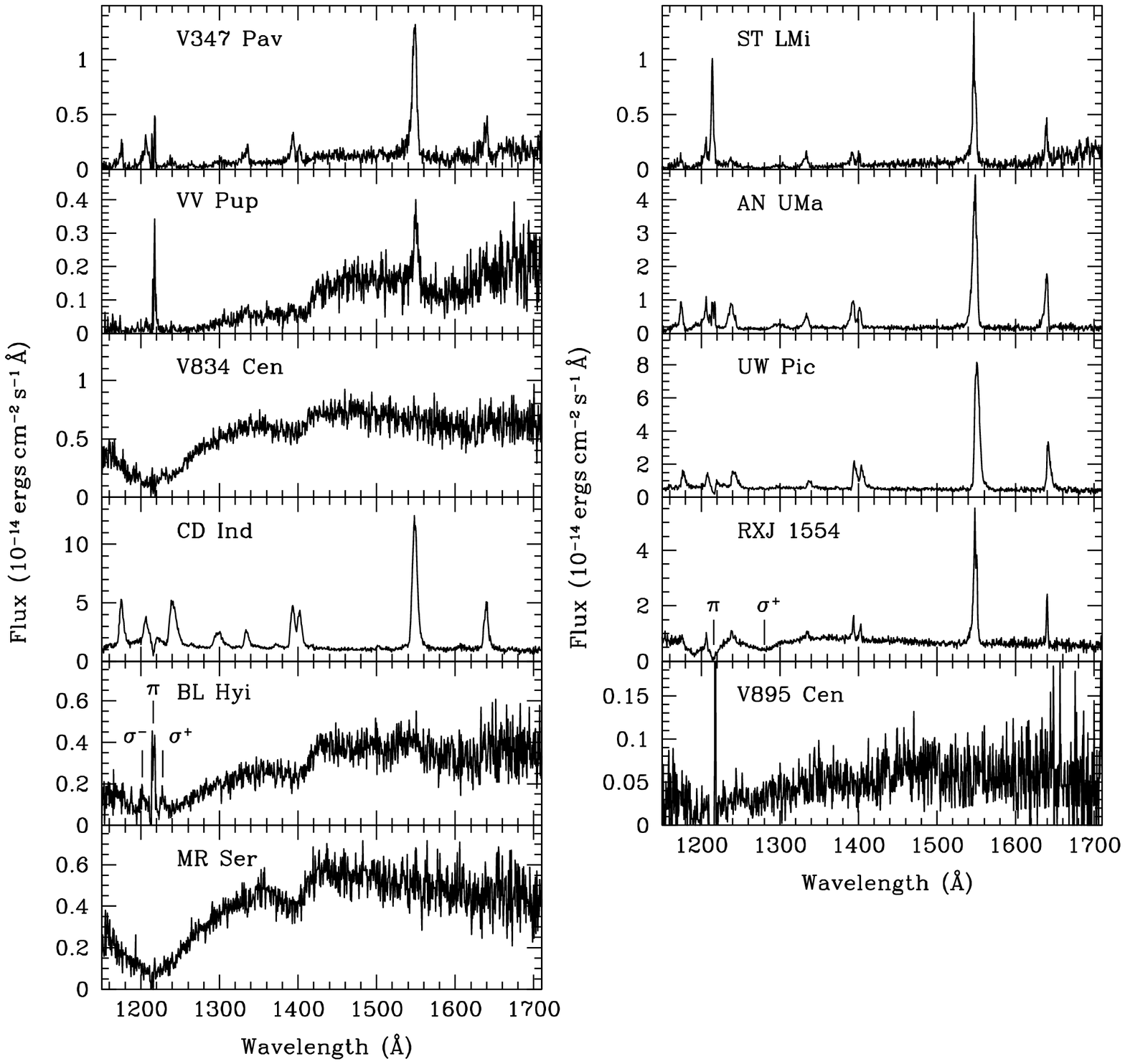}}
\caption[]{\label{f-all}Snapshot spectra of the group of eleven polars
  observed by \textit{HST}/STIS.}
\end{figure*}

FUV spectroscopy of eleven polars was obtained with \textit{HST}/STIS
in Cycle\,11 (Table\,\ref{t-log}). The data were obtained using the
G140L grating and the $52\arcsec\times0.2\arcsec$ aperture, providing
a spectral resolution of $R\approx1000$ over the wavelength range
$1150-1710$\,\AA. Since the total time involved in a snapshot
observation is short ($\sim35$\,min), we chose to make the
observations in ACCUM mode in order to minimize the instrument
overheads, resulting in exposure times of $\simeq12-15$\,min. As a
result of this choice, each observation resulted in a single time
averaged spectrum of each polar. For the analysis described here, all
of the data were processed within IRAF\footnote{IRAF is distributed by
the National Optical Astronomy Observatories, which is operated by the
Association of Universities for Research in Astronomy, Inc., under
contract with the National Science Foundation.} using CALSTIS
V2.13b. Figure\,\ref{f-all} shows the STIS spectra of the eleven
polars that were observed.

During target acquisition, \textit{HST} points at the the nominal
target coordinates and takes a $5\arcsec\times5\arcsec$ CCD image with
an exposure time of a few seconds. Subsequently, a small slew is
performed that centers the target in the acquisition box, and a second
CCD image is taken. The acquisition images for these observations were
obtained using the \FLP\ long-pass filter, which has a central
wavelength of 7228.5\,\AA\ and a full-width at half maximum (FWHM) of
2721.6\,\AA. We have made use of the STIS acquisition images to
establish the (quasi-simultaneous) brightness of the systems at the
time of the FUV observations. We computed instrumental magnitudes from
the acquisition images by performing aperture photometry with the
\texttt{sextractor} \citep{bertin+arnouts96-1}. We then obtained from
the \textit{HST} archive a number of acquisition images of
well-calibrated flux standards (GRW$+70^{\circ}$5824 and G191\,B2B) to
convert the instrumental magnitudes into \FLP\ magnitudes. Testing the
method on a number of STIS observations of objects of known and
constant magnitudes shows that the acquisition images can be used to
determine the visual magnitude to typically $\pm0.1$\,mag. The
response of the \FLP\ filter is closest to that of an $R$ filter, but
extends further both into the blue and red.

Based on the appearance of the FUV spectra (Fig.\,\ref{f-all}) and the
\FLP\ magnitudes (Table\,\ref{t-log}), three of the systems that we
observed were found in a high state: CD\,Ind, AN\,UMa and
UW\,Pic. Their FUV spectra are characterized by a nearly-flat
continuum superimposed by strong emission lines of He, C, N, and
Si. During the high state, the dominant FUV emission sources in polars
are the heated polar cap (or hot spot) and the accretion stream. Line
emission is the result of photoionization of the accretion stream by
the high-energy photons from shocks in the accretion column and/or the
impact regions of material on the WD surface.  CD\,Ind, AN\,UMa and
UW\,Pic exhibit similar emission line ratios of \CivL, \HEiiL, \NvL,
\SIivL, \CiiiLxi, \SIiiiLa\ and \CiiLxiii\ (see next section), the
latter being the weakest in all three systems (Fig.\,\ref{f-all}). The
\SIiiiL\ multiplet is also present in CD\,Ind and to some extent in
AN\,UMa but absent in the spectrum of UW\,Pic.  The emission lines are
highly asymmetric, consistent with their origin in the accretion
stream, which has a large velocity gradient. The narrow \La\
absorption (superposed on the broader emission) seen in the spectra of
CD\,Ind and UW\,Pic is of interstellar origin.

Eight polars were observed in a state of very low accretion activity,
as indicated by the weakness (V347\,Pav, VV\,Pup, ST\,LMi and
RXJ1554.2+2721,) or absence (V834\,Cen, BL\,Hyi, MR\,Ser and
V895\,Cen) of emission lines (Fig.\,\ref{f-all}). All of the low state
systems\footnote{Our definition of low state is mainly based on the
appearance of the FUV spectra and specifically on the direct detection
of the WD. This does not mean that the accretion activity has stopped
altogether but merely that it is low enough so that the WD stands out
over any other binary component} show the broad \La\ absorption
feature that is characteristic of a WD-dominated system. In addition,
the \FLP\ magnitudes of these systems are consistent with the
brightness level of previously observed low states
(Table\,\ref{t-log}). In V895\,Cen, the \FLP\ magnitude is close to
the observed high-state V magnitude. However, in this long period
polar the secondary star significantly contributes in the red also
during the high state \citep{craigetal96-1, stobieetal96-1,
howelletal97-2}. Consequently the high-to-low state variation in the
$R$ band is low and the \FLP\ magnitude is only of limited use to
determine the state of the system. Unfortunately, no $R$ low/high
state magnitudes have been published for V895\,Cen, except the USNO
catalogue entry of $R=$15.9. Comparing this value to \FLP=16.4
supports that the STIS observations were obtained during a low state
of accretion. The fact that the WD is clearly visible in these systems
allows us to obtain one of their fundamental system parameters, the WD
temperature (\Twd), as well as to estimate their distances.

\begin{figure*}
\centerline{\includegraphics[angle=0,width=14cm]{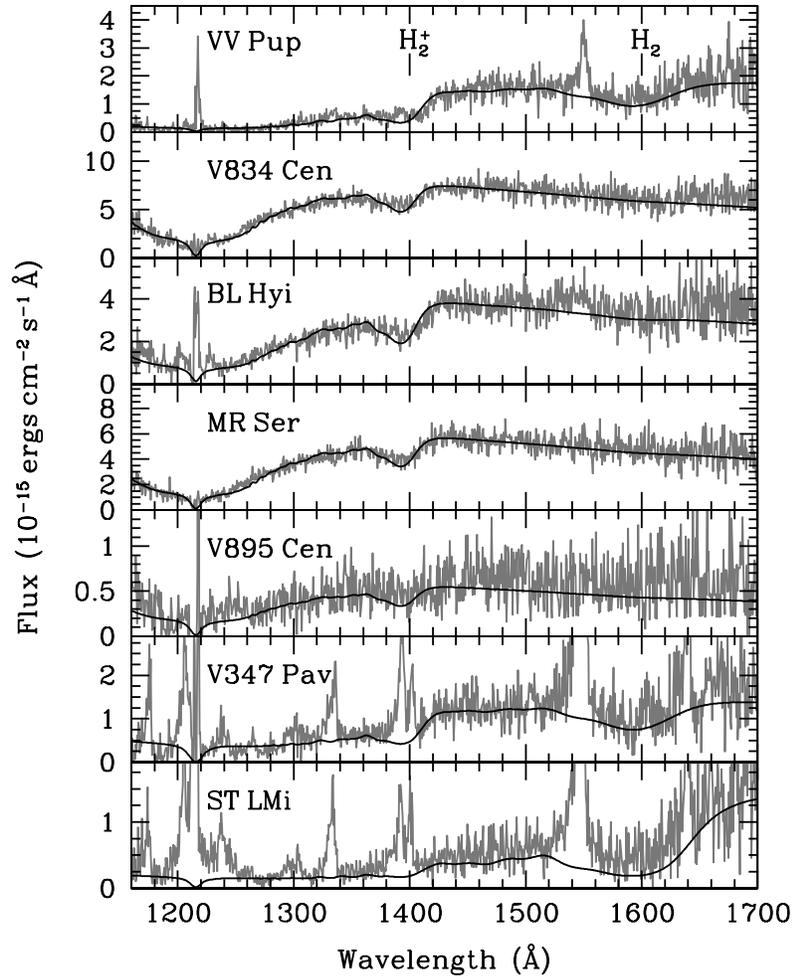}}
\caption[]{\label{f-low} \textit{HST}/STIS spectra (grey line) of the seven systems
found in a low-state of accretion. Plotted over (black line) is the
best 2-WD model for each of the systems. The parameter values for each
of the models are given in Table\,\ref{t-models}.}
\end{figure*}

A common feature in the FUV spectra of the low-state systems in
Fig.\,\ref{f-all} (see also Fig.\,\ref{f-low}), with the exception of
RX\,J1554.2+2721 (RX\,J1554 for simplicity), is the presence of
quasi-molecular hydrogen $\mathrm{H}^+_2$ at $\sim1400$\,\AA. Also
present in VV\,Pup, BL\,Hyi, V347\,Pav and ST\,LMi is absorption by
quasi-molecular hydrogen $\mathrm{H_2}$ at $\sim1600$\AA. The nature
of both quasi-molecular features, first identified in WDs by
\citet{koesteretal85-1} and \citet{nelan+wegner85-1}, is attributed to
perturbations of the potential energy between the ground and first
excited state of a neutral hydrogen atom caused by a proton or by
another neutral hydrogen atom. In both cases the perturbation results
in a shift of the energy of the emitted/absorbed photon, producing
$\rm H^+_2$ \La\ absorption at $\sim 1400$\,\AA\ in the case of a
perturbing proton, and producing $\rm H_2$ \La\ absorption at
$\sim1600$\,\AA\ in the case of a perturbing neutral hydrogen
atom. The strength of both features depends strongly on the
temperature of the hydrogen plasma (i.e. the WD atmosphere): the
$1400$\,\AA\ feature is present for $\Twd\lesssim 20\,000$\,K and the
1600\,\AA\ feature is only visible for $\Twd\lesssim 13\,500$\,K.

In contrast to the spectra of quiescent dwarf novae, the low state
spectra of polars show no metal absorption lines.  In fact, not a
single \textit{IUE} or \textit{HST} observations of any polar shows
metal absorption line strengths anywhere close to those observe in
non-magnetic systems \citep[see e.g.\,][]{gaensickeetal95-1}.  This is
in agreement with the idea that metal rich accreted material is
coupled to the magnetic field lines and sinks deep into the WD
atmosphere. Not until the gas pressure within the atmosphere exceeds
the magnetic pressure can the material break free and spread
laterally, and this occurs at large optical depths
\citep{beuermann+gaensicke03-2}.

An interesting feature in the low-state STIS spectrum of BL\,Hyi is
that it shows two emission lines centered on \La. We believe that this is
due to Zeeman splitting of \La\ and in Fig.\ref{f-all} we identify the
emission lines as the \La\ Zeeman $\pi$, $\sigma^{+}$ and $\sigma^{-}$
components. This is only the second time in which Zeeman splitting has
been observed in an emission line \citep[EF\,Eri;][]{seifertetal87-1}
and the first time it has been observed in \La. We defer further
discussion of this system to Sec.\,\ref{s-ind_systems}.

The FUV spectrum of RX\,J1554 reveals a very unusual broad absorption
line centered on $\sim 1280$\,\AA\ (Fig.\,\ref{f-all}). We identify
this feature as the \La\ Zeeman $\sigma^{+}$ component, split in a
field of $B>100$\,MG, analogous to that observed in the FUV spectrum
of AR\,UMa, whose magnetic field strength is $B=230\rm\,MG$ \citep[the
highest magnetic field among CVs; ][]{schmidtetal96-1,
gaensickeetal01-1}. A full account of RX\,J1554 is given by
\citet{gaensickeetal04-3}. Due to its high field strength, the
temperature determination of the WD in RX\,J1554 is subject to large
systematic uncertainties, and we will not further discuss the
properties of this object in the context of the present paper.

\section{Polars observed in a  high state}
\label{s-analysis}
\begin{figure*}
\centerline{\includegraphics[width=12cm]{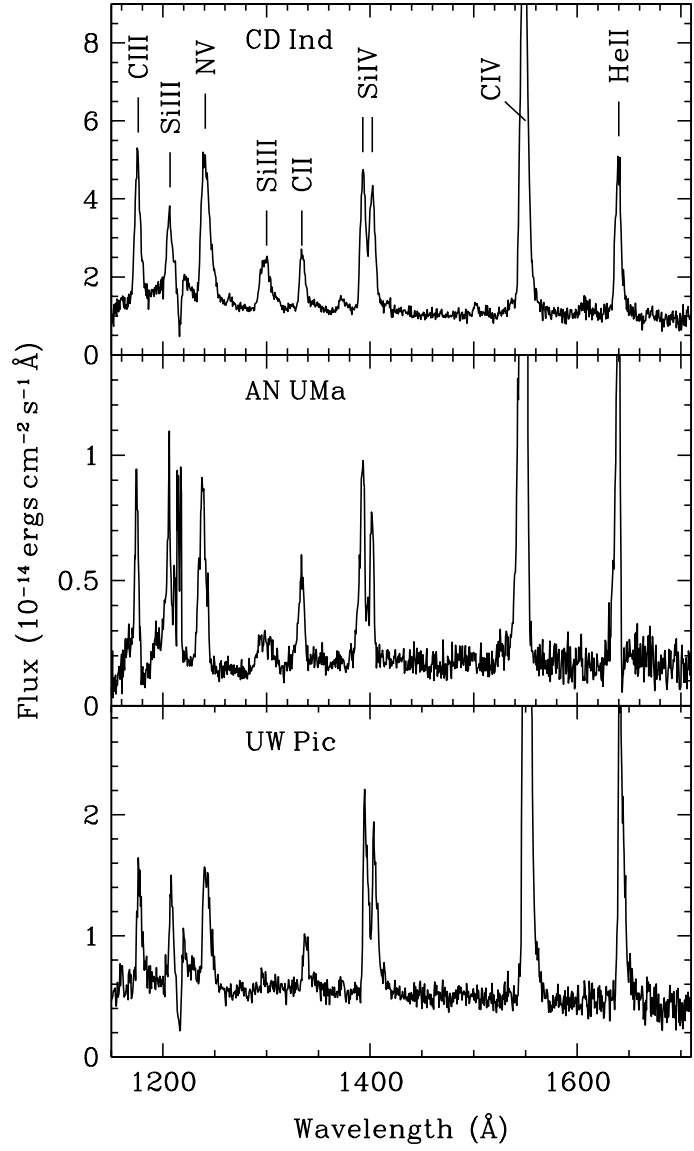}}
\caption[]{\label{f-high} Closer view of the \textit{HST}/STIS
  snapshot spectra of the three magnetic CVs found in a high state of
  accretion during the observations.}
\end{figure*}

In previous \textit{HST}/STIS and \textit{IUE} studies of the
prototype polar AM\,Her, \citet{gaensickeetal95-1, gaensickeetal98-2}
have shown that during high states a large heated pole cap dominates
the FUV continuum emission at wavelengths $\la1500$\,\AA. During the
high state, the emission from the polar cap in AM\,Her, even though
arising from a high-gravity photosphere, showed only a very shallow
broad \La\ absorption line. Using a semi-empirical model,
\citet{gaensickeetal98-2} showed that irradiation by thermal
bremsstrahlung and/or cyclotron radiation from the accretion column is
likely to flatten out the temperature gradient in the WD atmosphere
around the accretion column, resulting in a significantly decreased
depth of the \La\ absorption compared to an undisturbed WD
atmosphere. There is no evidence for broad \La\ absorption
($\simeq100$\,\AA) in any of the three high-state systems (see
Fig.\ref{f-high}). In contrast, in CD\,Ind (and, less clearly, in
UW\,Pic) a broad \textit{emission} bump centered on \La\ is observed.
Such feature could arise if irradiation results in a temperature
inversion in the WD atmosphere at a significant depth, where pressure
(Stark) broadening is sufficient to affect the \La\ profile over
$\simeq100$\,\AA. Full phase-resolved FUV spectroscopy would be
necessary to assess this hypothesis.

The line properties of the high-state systems in Fig.\,\ref{f-high}
are summarized in Table\,\ref{t-lines1}. A single Gaussian was fitted
to each of the lines in the normalized spectrum with the exception of
the \SIivL\ doublet where a double Gaussian fit was carried out. We
also computed the following line flux ratios: \HEiil/\Civl,
\SIivl/\Civl\ and \Nvl/\Civl\ (see Table\,\ref{t-ratios}), which fall
within the normal range observed in CVs
\citep{maucheetal97-1,gaensickeetal03-1}. As observed by
\citet{maucheetal97-1} from a small sample of polars, there is a
tendency within these three systems of exhibiting increasing
\Nvl/\Civl\ ratios with increasing \SIivl/\Civl, with the
\HEiil/\Civl\ ratio remaining approximately constant.

\begin{figure*}
\centerline{\includegraphics[angle=-90,width=12cm]{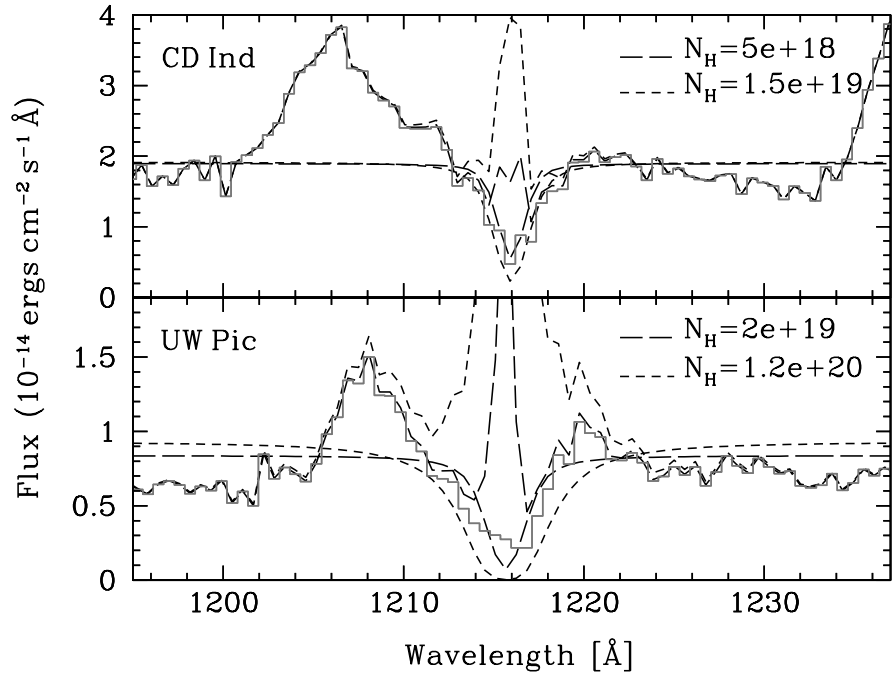}}
\caption[]{\label{f-is} Interstellar \La\ absorption of the spectra
of CD\,Ind (top panel) and UW\,Pic (bottom panel) plotted in grey. Dash lines
are the artificial \La\ profile for different values of $N_{\rm H}$ as
well as the corresponding corrected spectra.}
\end{figure*}

 In CD\,Ind and UW\,Pic, a narrow \La\ absorption line is detected,
which is of interstellar origin and allows a determination of the
column densities of neutral hydrogen along the line of sight to these
objects.  We modeled the \La\ interstellar absorption profile with a
Lorentzian profile \citep{bohlin75-1} with the neutral hydrogen column
density (\Nh) and the profile broadening as free parameters. The
broadening parameter used for both systems was the STIS resolution
(i.e.\,$\simeq$1.2\AA). The broadening produced by the turbulent
velocity of the interstellar medium is so small compared with the
instrumental resolution that it can be neglected here. In the case of
CD\,Ind, the resulting neutral column density is $(1.0\pm{0.5})\times
10^{19}\,\mathrm{cm^{-2}}$, well within the upper limit $\rm
4.5\times10^{20}\,cm^{-2}$ determined from radio 21\,cm data
(i.e. value of the entire content of $N_{\rm H}$ along the line of
sight of CD\,Ind; \citealt{dickey+lockman90-1}). Our estimate is
consistent with and should be more accurate than the more model
dependent value of $(0.4\pm1.0)\times10^{20}\,\mathrm{cm^{-2}}$
obtained by \citet{schwopeetal97-2} from ROSAT observations.  The top
panel of Fig.\,\ref{f-is} shows the computed interstellar \La\
profiles for values of $\Nh=0.5\times10^{19}\,\mathrm{cm^{-2}}$ and
$\Nh=1.5\times10^{19}\,\mathrm{cm^{-2}}$. Also shown is the FUV
spectrum of CD\,Ind corrected for the two respective values of
\Nh. For $\Nh>1.5\times10^{19}\,\mathrm{cm^{-2}}$, the correction
produces an unrealistically strong \La\ emission line. Due to clear
signs of \La\ emission in the spectrum of UW\,Pic, red-shifted due to
the orbital motion at the time of the STIS observations, fitting the
interstellar \La\ absorption is prone to some systematic
uncertainties, and we estimate a lower limit to be $\Nh >
2\times10^{19}\,\mathrm{cm^{-2}}$. This lower limit is consistent with
the value of $\Nh=1.2\times10^{20}\,\mathrm{cm^{-2}}$ reported by
\citet{reinschetal94-1}, which they derived from the analysis of
\textit{ROSAT} observations. Plotted on the bottom panel of
Fig.\,\ref{f-is} is the \La\ profile for values of
$\Nh=2.0\times10^{19}\,\mathrm{cm^{-2}}$ together with
$\Nh=1.2\times10^{20}\,\mathrm{cm^{-2}}$ of \citet{reinschetal94-1}.
As in the case of CD\,Ind, we corrected the observed spectrum for the
respective interstellar \La\ absorption profiles to be able to judge
the best models. Certainly, the value for $N_{\rm H}$ cannot be
greater than $1.2\times10^{20}\,\mathrm{cm^{-2}}$, as this would imply
a broader \La\ absorption profile than it is observed. Therefore, we
believe that the real value for $N_{\rm H}$ along the line of sight of
UW\,Pic is clearly in between the two values plotted on
Fig.\,\ref{f-is}, which are well within the upper limit
$4.5\times10^{20}\rm\,cm^{-2}$, estimated from radio 21\,cm data
\citep{dickey+lockman90-1}.

\section{Polars observed in a low state}
\label{s-analysis_low}
Previous analysis of \textit{IUE} and \textit{HST}/FOS observations of
AM\,Her \citep{gaensickeetal95-1, silberetal96-1}, DP\,Leo
\citep{stockmanetal94-1, schwopeetal02-1}, and HU\,Aqr
\citep{gaensicke98-2} have demonstrated that the FUV flux of these
polars during a low state can be interpreted in terms of photospheric
emission from the WD with two different temperature regions: a small
accretion-heated region~--~the polar cap (or hot spot)~--~ and a
larger area with a relatively low temperature representing the
``unheated'' WD photosphere.  The main focus of our analysis here is
to determine the temperatures of the WDs in seven of the polars
observed with STIS in a state of low accretion activity
(Table\,\ref{t-log}). The approach we have adopted is to fit the STIS
data of each of the systems with model spectra consisting of (a) a
single WD; (b) a WD and a power-law; and (c) a two-component WD
model. The two-component WD model is intended to represent the region
of the WD unaffected by accretion and the hot spot at the base of the
accretion column. We believe this model is the most physically
reasonable of the three approaches. However, the other two models are
important to assess the need for a second component and assess whether
we have statistically significant evidence for our specific hot spot
temperatures.

\subsection{WD model spectra and fitting routine}
We created a grid of local thermodynamic equilibrium (LTE), pure
hydrogen (DA) WD non-magnetic model spectra with temperatures ranging
from 8000\,K to 60\,000\,K, using the codes TLUSTY/ SYNSPEC
\citep{hubeny88-1, hubeny+lanz95-1}. The spectra were smoothed to the
resolution of the STIS. No additional broadening is expected since the
WDs in polars are phase-locked to the secondary and therefore
rotational effects are very small. We computed model spectra for three
values of the surface gravity, $\log g =$\,7.5, 8.0, and 8.5. In this
way we cover WD masses of $\Mwd\simeq 0.33-0.90\,\Msun$, which is the
typical range of WD masses found in CVs.

The assumption of LTE is well-justified for the temperature range
considered here. In contrast to non-magnetic CVs, the FUV spectra of
polars show no noticeable metal absorption lines, and therefore the
use of pure-hydrogen models is appropriate.  The field strengths of
the systems studied here (Table\,\ref{t-log}) are too low to cause
significant changes in the \La\ profile due to Zeeman-splitting. The
spectrum of BL\,Hyi \citep[$B=23$\,MG;][]{schwope+beuermann90-1,
ferrarioetal96-1} in Fig.\,\ref{f-all} shows Zeeman-splitting of \La\
in emission whose components are well within the broad \La\ absorption
originating in the WD atmosphere. This illustrates that fields of this
magnitude do not modify the \La\ profile significantly. Therefore,
despite the fact that we have not attempted to create spectra for
magnetic DAs, we believe the temperatures are reliable. In addition,
all the polars studied here are cold enough to show strong
quasi-molecular $\rm H_2^+$, and in some cases quasi-molecular $\rm
H_2$ absorption lines (see Fig.\,\ref{f-low}). The dominant
uncertainty in our approach is due to the unknown masses of the
WDs. In fact, \Twd\ and \Mwd\ (or $\log g$) are correlated as the
width of \La\ increases with higher WD mass as well as with lower
effective temperatures. We explore below quantitatively the dependence
of the best-fit \Twd\ on $\log g$.

Our fitting routine permits the use of single or multi-component
spectral models, and uses $\chi^2$ minimization as best-fit
criterion. In carrying out the fits, we masked the emission lines in
the spectra of VV\,Pup, BL\,Hyi, V895\,Cen, V347\,Pav and ST\,LMi; the
other low-state spectra were emission line-free. Free parameters in
the fit are the temperature (or power-law index) and the scaling
factor of each component. The scaling factor between the
model spectra and the observed data is defined as 
\begin{equation}
\label{normalization}
\frac{f}{H}=4\pi\frac{\Rwd^2}{d^2},
\end{equation}
where $f$ and $H$ are the ``de-reddened" flux and the Eddington flux
of the model, respectively, and \Rwd\ and $d$ are the WD radius and
the distance to the system. We have assumed negligible reddening for
all seven systems \citep[consistent with the absence of detectable
reddening in the \textit{IUE} spectra of V834\,Cen, BL\,Hyi, VV\,Pup
and MR\,Ser;][]{laDous91-1}. However, in order to test the effect of
reddening ($E_{(B-V)}$) on the fits we repeated the fits assuming
$E_{(B-V)}=0.05$, which can be considered a conservative upper limit
for these relatively nearby high-galactic latitude systems. The WD
temperatures did not change by more than $\simeq200$\,K. The scale
factors increased by a factor of $\simeq1.5$ compared with
$E_{(B-V)}=0$ because the de-reddened flux is higher. This translates
into a decrease of $\simeq20\%$ on the derived distance estimates,
which is within the error introduced by the uncertainty of the WD
mass (see Sect.\,\ref{s-distances}).

\subsection{WD effective temperatures}
As outlined above, our first approach was to fit the STIS spectra of
the seven low-state polars using a single temperature WD model. The
results from this exercise are reported in Table\,\ref{t-models}. The
uncertainties in the derived temperatures and scaling factors are
obtained from three separate fits using $\log g=7.5$, 8.0, and 8.5
models, and reflect the fact that the WD masses of these systems are
unknown. It is apparent that the uncertainty in the WD masses
translates into only a relatively small uncertainty in the derived
values for \Twd. The $\chi^2_{\nu}$ values obtained from the three
fits (i.e. for $\log g =$7.5, 8.0 and 8.5 respectively) to each of the
spectra are essentially the same. This underlines the fact that it is
not possible to independently determine $\Twd$ and $\Mwd$ from the
STIS data alone.

Guided by the experience of \citet{stockmanetal94-1},
\citet{gaensickeetal95-1}, \citet{gaensicke98-2} and
\citet{schwopeetal02-1}, as well as by the relatively high values of
$\chi^2_{\nu}$ for a number of systems (Table\,\ref{t-models}) we
fitted the STIS spectra of the seven polars with the two-component
models introduced above: a WD plus a power-law or two WD components
with different temperatures. In all cases the two-component model
improves the fit, most noticeably for V834\,Cen, BL\,Hyi and
MR\,Ser. However, based on $\chi^2_{\nu}$, we are unable to chose
between models that use a power-law or WD spectrum as the second
component. Furthermore, we found that the temperature of the second WD
component, typically in the range 30\,000\,K$-$70\,000\,K, was poorly
constrained by the fits. This is not very surprising since in our
spectral range, the slope of the hotter WD continuum is close to the
Rayleigh-Jeans limit and the \La\ profile is weak and relatively
narrow. Varying the temperature of the hot component had little effect
on the resulting temperature and scaling factor for the cold WD
component. Our conclusion is that while a second component improves
the fits, we cannot on the basis of statistics (a) distinguish between
power law fits and WD models for the second component, or (b)
accurately determine the temperature and fractional size of the second
component from the two WD fits.

We therefore decided to report the fits using two WD components with
the second component fixed at 50\,000\,K, comparable to the ``hot
spot'' temperature of DP\,Leo and AM\,Her derived by
\citet{stockmanetal94-1} and \citet{gaensickeetal98-2} from phase
resolved \textit{HST} data.  As for the single WD model approach, we
fitted the STIS spectra for three values of $\log g =$\,7.5, 8.0 and
8.5. These results are listed in Table\,\ref{t-models}, with the
errors reflecting the uncertainty in the masses of the WDs.  In all
cases the optical flux of the best two temperature fit model to each
of the FUV spectra is consistent with the simultaneous optical flux
obtained from the \textit{HST}/STIS acquisition image. The ratio of
the scale factors for a given distance and mass of the two temperature
components (Table\,\ref{t-models}; $F[\%]= N_2/N_1\times 100$) gives
an idea of the hot component contribution to our observations. This
value needs to be interpreted with caution as due to the nature of the
observations (snapshots which only cover a small portion of the
orbital cycle) we are unable to properly quantify the \textit{true}
hot spot surface area with respect to the entire WD. Overall, the
effective temperatures of the WDs did not change by large amounts
compared to the single WD fits, but by analogy to the few polars for
which phase-resolved low-state FUV observations exist, we believe that
the WD temperatures determined from the two-component model are
preferable.

\subsection{Distances}
\label{s-distances}
Even though accurate distances to a few CVs have recently been
determined directly through ground and space based parallax programs
\citep{harrisonetal00-1, thorstensen03-1, beuermannetal03-1,
beuermannetal04-1}, the bulk of all available distance estimates is
still based either on the absolute magnitude in outburst-orbital
period relationship found by \citet{warner87-1} (only in the case of
dwarf novae), or on red/infrared observations of the donor stars in
these systems. The latter technique is based primarily on the
quasi-independent relation between surface brightness of cool stars in
the $K$-band and their temperature (or luminosity) empirically found
by \citet{bailey81-1}, and later refined by \citet{ramseyer94-1}. The
most crucial assumption of this method is that the observed $K$-band
magnitude is entirely due to emission from the secondary.  Whereas in
dwarf novae the outer (cooler) regions of accretion disks may
contribute to the observed red/infrared flux the main source of
contamination in polars is cyclotron emission. Hence, the $K$-band
magnitude method is prone to underestimate the distances. A
significant improvement of this method is achieved if the secondary
can be detected spectroscopically; \citet{beuermann+weichhold99-1}
presented a calibration of the surface brightness in the TiO
absorption bands prominent in M-dwarfs found in short-period CVs, and
conclude that with the appropriate data, distance estimates with an
accuracy of $10\%$ can be obtained.

Here we make use of an independent method to estimate the distances of
the polars observed in the low state, based on the surface brightness
of the primaries in these systems through the normalization factor
defined in Eqn.\,\ref{normalization}. To make use of this relation, we
need to assume a mass-radius relation to convert the WD mass/$\log g$
into a stellar radius. For the low temperatures of the WDs considered
here, the Hamada--Salpeter mass-radius relation
\citep{hamada+salpeter61-1} is accurate enough. Obviously, the unknown
WD mass is the largest systematic uncertainty in these estimates. Our
WD model fits primarily reflect the shape of \La. At a given
temperature, the \La\ profile increases in width as the mass
increases. At a given mass, the \La\ profile decreases in width as the
temperature increases. The physical reasons for these dependencies are
as follows: A higher WD mass (higher $\log g$) implies a higher
density in the WD atmosphere, resulting in an increase in the width of
\La\ due to the stronger Stark broadening. A higher temperature
results in a narrower \La\ profile, as the number of neutral hydrogen
atoms decreases with increasing \Twd. The same observed \La\ profile
can hence be fit approximately well with a somewhat hotter high-mass
WD or a somewhat colder low-mass WD. Furthermore, the surface area of
the WD is $\propto\Rwd^2$, and $\Rwd$ is anticorrelated with
$\Mwd$. Combining all effects, we find that the systematic uncertainty
in the distance is $\pm13-23$\,\% for WD masses in the range
$\Mwd=0.33-0.90$\,\Msun\ (see Table\,\ref{t-distances}). Using the WD
as an independent distance estimator is not only useful to confirm
distance measurements based on \citet{bailey81-1}'s method (see
e.g. \citealt{gaensickeetal99-1, araujo-betancoretal03-1,
hoardetal04-1}), but is the only feasible way to obtain distance
estimates for systems in which the secondary is not detected, such as
those likely to contain a brown dwarf donor (e.g.  EF\,Eri:
\citealt{beuermannetal00-1}, HS\,2331+3909:
\citealt{araujo-betancoretal04-2}).

\subsection{\label{s-ind_systems} Individual systems}
Here we discuss the results for the individual systems and compare our
findings to previous studies. 

\textit{VV\,Pup~---} \citet{liebertetal78-1} fitted a blackbody to an
optical low state spectrum and estimated $\Twd\simeq9000$\,K and
80\,pc$<d<$150\,pc. \citet{bailey81-1} estimated $d\simeq144$\,pc
based on the low-state $K$ band magnitude measured by
\citet{szkody+caps80-1}. VV\,Pup has been considered to host the
coldest WD in a CV, however, our value of $T_{\rm wd}= 11\,900$\,K
moves it to a significantly higher temperature. Blackbodies are
relatively crude approximations to a WD spectra, and our STIS result
should supersede the older estimate. Our estimate of the distance to
VV\,Pup, $d=151\pm{28\atop34}$\,pc, agrees well within the errors with
that obtained by \citet{bailey81-1} using Bailey's method, which in
general gives lower limits and therefore we believe our value to be
more robust.

 
\textit{V834\,Cen~---} \citet{maraschietal84-1}'s blackbody fit to
simultaneous optical and \textit{IUE} low-state spectra gave a
$\Twd=26\,500$\,K WD within a distance of $d=50-200$\,pc from the low
state $R$ and $I$ magnitudes. They noted, however, that the WD
interpretation of the UV continuum was somewhat ambiguous, as no broad
\La\ absorption was observed. \citet{cropper90-1} applied Bailey's
method using $K=13.1$, as reported by \citet{maraschietal84-1} during
a high state. This value of $K$ is clearly an upper limit to the
$K$-brightness of the secondary in V834\,Cen and therefore the derived
distance $d = 86$\,pc is an absolute lower limit. Further optical
low-state spectroscopy and photometry was reported by
\citet{puchnarewiczetal90-1}, \citet{schwope90-1} and
\citet{ferrarioetal92-1}. \citet{puchnarewiczetal90-1} obtained
$\Twd=12\,000$\,K from a blackbody fit to the optical low-state
continuum and $d>70$\,pc from the spectral signature of the companion.
Interestingly, both \citet{schwope90-1} and \citet{ferrarioetal92-1}
suggested a two-temperature model for the WD to fit the observations.
\citet{ferrarioetal92-1} showed a very smooth, quasisinusoidal blue
light curve (their Fig.\,5), which they interpreted as a hot spot of
$T_{\rm spot} \simeq 50\,000$\,K on a $\Twd=15\,000 - 20\,000$\,K WD
with probably some cyclotron emission from a residual accretion
column. \citet{schwope90-1} modeled phase-resolved low state optical
spectroscopy with a $\Twd=15\,000$\,K WD and a spot of
30\,000\,K. More recently, \citet{gaensicke98-1} reached the same
conclusion as \citet{ferrarioetal92-1} and \citet{schwope90-1} about
the need of two-temperature model to fit the observations of
V834\,Cen. They used \textit{IUE} low state spectroscopy and find
$\Twd=15\,000$\,K and $\Tspot=30\,000$\,K. Our STIS value for the WD
temperature of V834\,Cen, $\Twd=14\,300$\,K, confirms the values
reported by \citet{schwope90-1} and \citet{gaensicke98-1}, and our
data clearly underline the need of a two-component model for V834\,Cen
in the low state. Our distance estimate, $d = 144\pm{18\atop23}$\,pc
appears realistic. 


\textit{BL\,Hyi~---} As already mentioned in
Sec.\,\ref{s-observations}, the spectrum of BL\,Hyi exhibits
Zeeman-splitting in the \La\ emission (see Fig.\,\ref{f-all}). We
measure the spacing between the central $\pi$ component and the
$\sigma^-$ and $\sigma^+$ components of the \La\ emission
Zeeman-splitting to be $\Delta\lambda\sim14$\,\AA. Using
$\Delta\lambda = e\lambda^2 B/4\pi m_ec^2$ (where $e$ is the electron
charge and $m_e$ is its mass), or $B = 1.45\Delta\lambda$\,[MG] (for
values of $\Delta\lambda$ in \AA), we infer a field strength of $B\sim
20$\,MG which is consistent with the value measured from cyclotron
radiation and Zeeman absorption features
\citep[$B=23$\,MG;][]{schwope+beuermann90-1,
ferrarioetal96-1}. Considering that $B\propto 1/r^{3}$, the \La\
emission line must arise therefore from emission regions very close to
the white dwarf surface. A previous study of BL\,Hyi reported by
\citet{wickramasingheetal84-1} gave a value for the temperature of the
WD of $\Twd\simeq20\,000$\,K. In this case, the authors modeled
low-state optical spectroscopy with simple magnetic WD models.  Our
estimate, $\Twd = 13\,300$\,K, is significantly different from the
above but agrees well with the estimate of \citet{schwopeetal95-1},
$\Twd = 13\,000$\,K, obtained from modeling optical low-state
spectroscopy of BL\,Hyi. Here the authors used a distance to the
system of $d=132\pm20$\,pc, obtained via Bailey's method.  Our
distance estimation $d = 163\pm{18\atop26}$\,pc is consistent with the
above.


\textit{MR\,Ser~---}\citet{mukai+charles86-1} fitted optical low-state
spectroscopy with a composite model consisting of a blackbody plus
M-dwarf template and found $\Twd\simeq9000$\,K and $d=142$\,pc. They
mentioned, however, that the choice of \Twd\ was somewhat arbitrary, and
they were mainly interested in the secondary star. \citet{szkody88-1}
fitted \textit{IUE} intermediate state data with WD models and found
$\Twd\simeq20\,000\,K$. The absence of a \La\ absorption line,
characteristic of the WD photosphere, casts some doubt on the
interpretation of the authors and the inferred
value. \citet{schwopeetal93-1} estimated $d=139\pm13$\,pc from the
detection of the M5--M6 secondary star in optical low state
spectroscopy which agrees well with our estimate, $d =
160\pm{18\atop26}$\,pc.  The same authors derived
$\Twd=8500-10\,000$\,K from modeling the low state spectrum with WD
spectra. The value obtained in the present work for the temperature of
the WD, $T_{\rm wd}=14\,200$\,K, is significantly higher than the
previous estimates, but being based on the unambiguous signature of the
WD in the FUV, we believe that our value is more accurate.


\textit{V895\,Cen~---} \citet{howelletal97-2} estimated the distance to
V895\,Cen to be $d\simeq250-295$\,pc by using the absolute visual
magnitude of the best main-sequence template fit to the optical
spectrum and the measured $V$-magnitude. Our estimate of $d=
511\pm{60\atop81}$\,pc puts V895\,Cen significantly further away. As in
the case of Bailey's method, the method employed by
\citet{howelletal97-2} gives lower limits for the distance since other
sources within the system (apart from the secondary) might have contributed
to the measured visual magnitude.


\textit{V347\,Pav~---} \citet{ramsayetal04-1} modeled UV broadband
fluxes obtained from the Optical Monitor (OM) on board of the
\textit{XMM}-Newton observatory to estimate $\Twd<10\,200$\,K and
$d\simeq40-50$\,pc. Their observations were obtained during a high
state, so the contribution of the accretion spot/stream to the UV flux
was likely to have contaminated the emission from the WD, and broad-band
fluxes were probably insufficient to permit a reliable determination of
$\Twd$. A WD temperature of $\Twd = 11\,800$\,K at a distance
$d=171\pm{33\atop38}$ determined from our detailed STIS spectroscopy is
certainly a more robust result. With Ramsay's
\citeyear{ramsayetal04-1} distance of $\simeq40-50$\,pc, V347\,Pav
would be one of the closest CVs known, and a ground-based parallax
program would be worthwhile to test this hypothesis.


\textit{ST\,LMi~---}The first report on the temperature of the WD in
ST\,LMi was made by \citet{schmidtetal83-1}. In their paper they
estimated $\Twd=10\,000-25\,000$\,K by comparing a low state optical
spectrum of ST\,LMi to that of AM Her. \citet{szkodyetal85-1} obtained
a WD temperature of $\Twd=13\,400\pm2000$\,K from blackbody fits to
\textit{IUE} low state spectra and optical/IR
photometry. \citet{baileyetal85-1} estimated $d=136$\,pc based on the
faint-phase $K$ magnitude, and $\Twd=12\,000-30\,000$\,K from $U-B$
and $B-V$ colours. \citet{mukai+charles87-1} modeled optical low-state
spectroscopy of ST\,LMi with a composite model consisting of a
blackbody plus M-dwarf template (M5--M6). They found that any of the
blackbody curves with temperature values within the range given by
\citet{baileyetal85-1} (i.e. $\Twd=12\,000-30\,000$\,K) could equally
well fit the observations. Our value of $\Twd=10\,800$\,K is the first
unambiguous measurement of the WD temperature in ST\,LMi. Our distance
estimate of $d=115\pm{21\atop22}$\,pc is consistent with Bailey's
\citeyear{baileyetal85-1} value of $d=136$\,pc.

\section{\label{s-discussion}Discussion}

We have obtained \textit{HST}/STIS FUV observations of a group of eleven
polars, eight of which were found in a low accretion state.
Thanks to the reduction of the mass transfer from the secondary, we
were able to have a direct view at the WD allowing us to measure its
temperature and to estimate the distances to the systems.

\subsection{\label{s-wd_temperatures}WD temperatures in the context of
  CV evolution} Measuring WD effective temperatures for seven polars
from high-quality STIS data represents a substantial addition to the
number of polars with well-determined WD temperatures. In order to
discuss our results in the general context of CV\,WD temperatures, we
have compiled in Table\,\ref{t-temperatures} a list of reliable CV\,WD
temperature measurements. We only include values derived from WD model
fits to FUV spectra which clearly show \La. The only exception is the
short-period polar EF\,Eri, in which the WD is unambiguously detected
at optical wavelength during a low state \citep{beuermannetal00-1}. We
also exclude the WDs in the recently discovered polars,
SDSSJ155331.12+551614.5 and SDSSJ132411.57+032050.5
\citep{szkodyetal04-2}, and in HS\,0922+1333 \citep{reimers+hagen00-1}
and WX\,LMi=HS\,1023+3900 \citep{reimersetal99-1}, as so far there is
no evidence that they have started Roche lobe overflow mass transfer.

\begin{figure}
\centerline{\includegraphics[angle=270,width=10cm]{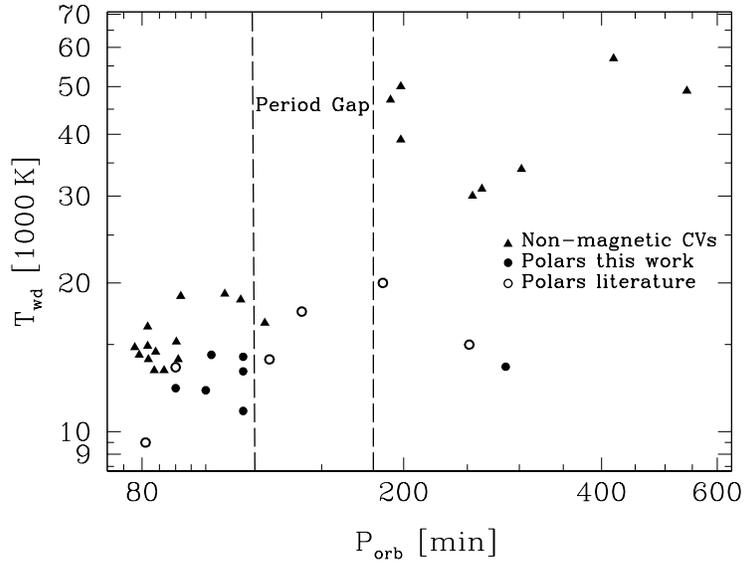}}
\caption[]{\label{f-t_porb} WD temperatures versus orbital
  period. Filled triangles are non-magnetic CVs, filled circles are
  those low-state polars under study in this article and open circles
  are polars not contained here.}
\end{figure}

Figure\,\ref{f-t_porb} shows the CV\,WD \Twd\ as a function of the
orbital period for magnetic and non-magnetic CV\,WDs.  Two important
results are evident: (a) \Twd\ decreases toward shorter periods and
(b) magnetic CVs show lower \Twd\ than non-magnetic CVs at any given
period. The new results confirm suggestions by \citet{sion91-1}, who
used the best available temperature information at the time: only two
\Twd\ measurements were reliable in the sense that we have defined
here (AM Her: \citealt{heise+verbunt88-1} and V834\,Cen:
\citealt{schwope90-1}).

The evolution of single WDs is driven by the cooling of their
degenerate cores. Isolated WDs reach photospheric temperatures of
$\Twd\thickapprox 8000$\,K after $\simeq1$\,Gyr and
$\Twd\thickapprox4500-5000$\,K after $\simeq4$\,Gyr
\citep{salarisetal00-1, wood95-1}. CVs are believed to spend on
average $\simeq2$\,Gyr as detached WD/main sequence binaries
\citep{schreiber+gaensicke03-1}. Once they start mass transfer, they
evolve above the gap on time scales of a few 100\,Myr, and below the
gap on time scales of several Gyr \citep{kolb+stehle96-1}. Comparing
these time scales and the observed \Twd\ of CV\,WDs to the theory of
WD cooling, it is evident that long term accretion-induced heating
(with possibly some additional heating occurring during nova
outbursts) is essential to compensate the secular core cooling.

According to the standard model of CV evolution \citep[e.g.][ for a
review]{king88-1}, systems evolve to shorter orbital periods
responding to angular momentum loss (AML) from the binary, with
magnetic braking dominating in systems above the period gap and
gravitational radiation in those below. While the standard theory
explains the period gap by invoking a sudden decrease in magnetic
braking (= decrease in accretion rate), it fails to explain a number
of features that characterize the orbital period distribution of CVs
\citep[e.g.][]{kolb+baraffe99-1, patterson98-1, kingetal02-1}.
Several authors \citep[][ among others]{patterson98-1, kingetal02-1}
have suggested an additional AML mechanism working below the period
gap to resolve some of the discrepancies between the theory and the
observations; notably the observed minimum period ($P_{\rm
min}\sim 80$\,min) compared to the predicted value ($P_{\rm min}\sim
65$\,min), and the missing number of observed systems around $P_{\rm
min}$, and below the 2\,h orbital period in general, predicted by
population synthesis models
\citep{kolb+baraffe99-1}. \citet{townsley+bildsten03-1} derived a
relation between the mean mass accretion rate
($\langle\dot{M}\rangle$) and temperature of the WD so that for a
given temperature $\langle\dot{M}\rangle$ can be estimated.  Using
their $\langle\dot{M}\rangle$-$\Twd$ relations,
\citet{townsley+bildsten03-1} concluded that the temperatures of
non-magnetic CVs below the gap revealed mass accretion rates higher
than those computed from gravitational radiation alone, providing an
indirect evidence for an additional AML mechanism below the gap, which
increases $\langle\dot{M}\rangle$.

Whereas the standard scenario predicts that magnetic and non-magnetic
CVs will evolve in an indistinguishable way,
\citet{wickramasinghe+wu94-1} and \citet{lietal94-1} have predicted
that the strong magnetic fields in polars would result in the
formation of closed field lines between the WD and the donor star. This will
effectively reduce the number of the open field lines responsible
for the drain of angular momentum from the system. Consequently, the
mass transfer rates in polars are expected to be lower, and their
evolution time scales longer compared to non-magnetic CVs.

Fig.\,\ref{f-t_porb} shows that for any given orbital period, WDs in
magnetic CVs are colder than in non-magnetic CVs. Following the work
of \citet{townsley+bildsten02-1, townsley+bildsten03-1,
townsley+bildsten04-1} this implies \textit{lower} secular accretion
rates in the strongly magnetic polars compared to the non-magnetic
dwarf novae, and is consistent with the hypothesis of reduced magnetic
braking in polars. The difference in WD effective temperature between
magnetic and non-magnetic CVs is largest above the gap, but it is
still significant below the gap. Assuming gravitational radiation as
the only AML mechanism below the gap, there should be no differences
between the WD effective temperatures of magnetic and non-magnetic
CVs. Our finding of lower temperatures in polars compared to
non-magnetic CVs also below the gap can be understood in a scenario
where residual magnetic braking is important below the gap, but is
reduced by the presence of a strong magnetic field on the WD.

While a difference in the long-term accretion rate between magnetic
and non-magnetic CVs is the most likely mechanism to explain the
differences in their WDs effective temperature, several other physical
factors that could partially account for this effect need to be
considered.  As discussed by \citet{townsley+bildsten03-1} \Twd\ is a
function of both, the secular mean accretion rate and the WD mass, as
a larger mass implies a higher accretion luminosity per accreted gram
of material. Considering this degeneracy in ($\dot M, \Mwd$), the
lower temperatures found in polars could be explained within a
scenario where magnetic and non-magnetic CVs have the same mean
accretion rates, but magnetic CVs have on average less massive WDs
than non-magnetic CVs. While for single WDs \textit{the contrary is
true}, i.e. magnetic WDs are more massive than non-magnetic ones
\citep{liebertetal03-1}, the situation is much less clear for CVs, as
the CV\,WD mass values (with just a few exceptions) are subject to
extreme systematic uncertainties.

Another effect to consider is the degree to which short term
variations in the accretion history of the CVs being compared
compromise the interpretation of \Twd\ as indicative of the long-term
($10^5$\,yr) accretion rates that control the evolutionary history of
CVs. Observations of some dwarf novae, e.g VW\,Hyi
\citep{gaensicke+beuermann96-1} indicate that a WD returns to a
quiescent state within 20 days of outburst, while the WD in WZ\,Sge
had not fully cooled to its pre-outburst temperature 18 months after
outburst \citep{longetal03-1, sionetal03-2}.  These differences likely
reflect the different amounts of mass accreted during an outburst,
much more in WZ\,Sge than in VW\,Hyi.  The situation in polars is less
understood, since essentially only one intensive study of
interoutburst temperature measurements has been conducted. In AM\,Her,
\citet{gaensickeetal95-1} have analyzed all available low-state
\textit{IUE} observations of AM\,Her, and found consistently
$\Twd\simeq20\,000$\,K, independent of the time passed since the
previous high state. However, given the fact that the temperatures
listed in Table\,\ref{t-temperatures} and shown in
Fig.\,\ref{f-t_porb} were measured at random inter-outburst phases for
the non-magnetic systems, and at random points during low states for
the polars, some conspiracy appears to be necessary to explain the
observed differences in \Twd\ by the effect of short-term
heating/cooling. Obviously, given that the difference in \Twd\ between
magnetics and non-magnetics is smallest in the region below the gap,
this is where it is most difficult to rule out the effects of short
term heating, and that only few systems with reliable \Twd\ are
available above the gap, additional data would be desirable to improve
the statistics of the temperature comparison.

\subsection{Low-states versus high-states}
In our survey of eleven polars, we have observed eight systems in a
low-state of accretion (or $73\%$). Considering the relatively small
number of systems, the percentage of those found in low state is in
good agreement with the numbers found by \citet{ramsayetal04-1}.  In
their study they observed a total of 37 polars with
\textit{XMM-Newton} of which 16 ($43\%$) were found in a
low-state. They also re-examine a sample of the \textit{ROSAT} All-Sky
Survey data and found that 16 out of the 28 polars that were
\textit{not} discovered by \textit{ROSAT} ($57\%$) were in a
low-state. In their study the authors concluded that there was no
evidence for a correlation between the orbital period and the
occurrence of low states. That means that low states appear not to be
connected with the evolution of the systems. Dedicated monitoring
programs, aiming at mapping out the frequency of high and low states
in polars, is needed in order to confirm this point and to search for
any correlation with the WD magnetic field strength, which will
improve our understanding of CV evolution.

\subsection{Space density}
Standard population models \citep[e.g.][]{dekool92-1, kolb93-1,
politano96-1} typically give a galactic CV space density of up to $
10^{-4}\,\mathrm{pc}^{-3}$. Assuming a scale height of $\sim150$\,pc
above the galactic plane \citep{patterson84-1,thomas+beuermann98-1}, a
mid-plane density of $10^{-4}\,\mathrm{pc}^{-3}$ corresponds to
$\simeq680$ CVs within a distance of 150\,pc (accounting for the
exponential drop-off in density over one scale height decreases the
effective volume by a factor $\simeq2$)~--~compared to a few tens of
known CVs with a confirmed $d\la150$\,pc.

Within the error bars, we find six polars in our STIS sample that are
within a distance of $\simeq150$\,pc: VV\,Pup, V834\,Cen, BL\,Hyi,
MR\,Ser, V347\,Pav and ST\,LMi. Compiling the distance estimates for
the other known polars from the literature, three additional systems
are within a distance of 150\,pc: AM\,Her ($d=79$\,pc,
\citealt{thorstensen03-1}), EF\,Eri ($d\sim150$\,pc,
\citealt{beuermann00-1, thorstensen03-1}), and AR\,UMa
($d\simeq88$\,pc, \citealt{remillardetal94-2}). Again assuming a scale
height of 150\,pc, the effective volume sampled is
$6.81\times10^6\mathrm{pc^3}$, which implies a space density of
\textit{known} polars of $1.3\times10^{-6}\mathrm{pc^{-3}}$, compared
to $\sim 4.4\times10^{-6}\mathrm{pc^{-3}}$ of the \textit {known}
non-magnetic CVs with $d\la150$\,pc.

Considering that the majority of polars have been discovered by their
X-ray emission, that \textit{ROSAT} has been the only all-sky-survey
in this wavelength band, that polars spend on average $\simeq50$\% of
their time in a low state, and that the local interstellar medium is
inhomogeneous \citep{warwicketal93-1} (which favors the discovery of
systems in certain parts of the sky more than others), it is clear
that the number derived above is an absolute lower limit on the space
density of polars, and that the true value could be easily a
factor $\simeq2$ higher.

The total ratio of known magnetic/non-magnetic CVs is $\sim 22\%$
 \citep[if we only include confirmed AM\,Her stars and Intermediate
 polars;][]{downesetal01-1}\footnote{Alternatively, if we only
 consider those systems with $d\la150$\,pc, then the ratio of
 magnetic/non-magnetic CVs is $\sim 40\%$ (including the corresponding
 intermediate polars to the list of polars to account for the magnetic
 CVs)} whereas it is $\sim5\%$ for isolated WDs
 \citep{jordan97-1,wickramasinghe+ferrario00-1}. If we assume that the
 fractional birthrate of magnetic to non-magnetic WDs should be
 similar for single WDs and those in CVs, then we must think of a
 mechanism that increases the fraction of magnetic CVs. The first and
 obvious mechanism are observational selection effects. Whereas it is
 true that \textit{ROSAT} has been very efficient in finding magnetic
 CVs, the same is true for amateurs finding dwarf novae, and a
 quantitative comparison is difficult. However, spectroscopic optical
 surveys that are not biased towards X-ray emission but discover CVs
 on the base of their colours/emission lines consistently report the
 fraction of magnetic CVs in the range of $\simeq10-20$\,\%
 \citet{gaensickeetal02-1, marshetal02-1, szkodyetal02-2,
 szkodyetal03-2, szkodyetal04-1}.  An \textit{intrinsic} mechanism of
 increasing the fraction of magnetic CVs is inherent to the discussion
 in Sect.\,\ref{s-wd_temperatures}: if strong magnetic fields on the
 WD indeed inhibit the efficiency of magnetic braking, then the
 evolution time scales of magnetic CVs are correspondingly longer, and
 the likelihood of discovering them at a given orbital period is
 increased with respect to non-magnetic CVs. A possible caveat to this
 line of reasoning is that the true fraction of isolated
 magnetic/non-magnetic WDs might be higher than suggested by previous
 studies. \citet{liebertetal03-1} have recently shown that the
 fraction of isolated magnetic/non-magnetic WDs of the local sample,
 which is considered as complete (to distances of 13\,pc), is
 $11\%\pm5\%$ for $B>2$\,MG.  This value is marginally consistent with
 the $\sim5\%$ quoted by \citet{jordan97-1} and
 \citet{wickramasinghe+ferrario00-1}, but it underlines the need to
 understand observational selection effects.

\section{Conclusions}
We studied a group of eleven polars observed with \textit{HST}/STIS as
part of our snapshot program aimed at compiling a homogeneous data
base of high-quality FUV spectra of CVs. Of the eleven AM Her systems
under study here, we found eight in a low-state of accretion. One
described in detail elsewhere \citep{gaensickeetal04-3} is RXJ1554
with $B>100$\,MG, the third high-field polar.  A second is BL\,Hyi,
for which we confirm a field of $\sim$20\,MG from Zeeman splitting of
\La\ in emission, the first time \La\ Zeeman splitting has been
detected in emission. Thanks to the reduction of the mass accretion
rate during low-states a direct study of the WD in these systems has
been possible. We have derived their temperatures and distances to the
system of the low-state polars in an uniform way by fitting their FUV
spectra with WD models (with the exception of RX\,J1554, where the
high field strength impedes an accurate \Twd\ determination).  Our
main conclusions are as follows:

\begin{enumerate}

\item The low-state spectra of VV\,Pup, V834\,Cen, BL\,Hyi, MR\,Ser,
894\,Cen, V347\,Pav, and ST\,LMi can be well described with 
two-temperature models: a hot component with a small area
(a hot polar cap) and a cold component which is ascribed to the
`unheated' photosphere of the WD. The seven polars analyzed have WD
temperatures in the range 10\,800\,K--14\,200\,K.

\item Our analysis increases the sample of polars with well-determined
\Twd\ from 6 to 13. Combining the $\Twd$ obtained here with all
reliable CV\,WD temperatures from the literature we find that the WDs in
magnetic CVs are colder than those in non-magnetic CVs at any given
orbital period.  

\item Using the $\Twd-\langle\dot{M}\rangle$ relation derived by
\citet{townsley+bildsten03-1} we conclude that the temperatures of
polars below the gap are consistent with those expected from
gravitational radiation as the only mechanism removing angular
momentum from the systems. This is in contrast with the result for
non-magnetic CVs below the gap obtained by
\citet{townsley+bildsten03-1}, in which they found that an extra
angular momentum loss mechanism is needed to explain the measured
accretion rates. Above the gap, the WD temperatures in polars imply
lower mean accretion rates than predicted by magnetic
braking, which supports the hypothesis that a strong magnetic field on
the primary will reduce the efficiency of magnetic braking. 

\item The total ratio of known magnetic/non-magnetic CVs is $\sim
22\%$ compared to $\sim 5\%$ observed in isolated WDs. Reduced
magnetic braking in polars would increase the time scale of their
evolution, which might explain their apparent overabundance compared
to non-magnetic CVs.

\item We have obtained relatively accurate distance estimates to the
low state systems from the WD surface brightness. Using these
distances and those published in the literature, we find a lower limit
to the space density of polars of $1.3\times10^{-6}\,\rm
pc^{-3}$. 

\end{enumerate}

\acknowledgements We thank both, the anonymous referee and the editor,
James W. Liebert, for helpful comments and suggestions. Support for
this work was provided by NASA through grants GO-9357 and GO-9724 from
the Space Telescope Science Institute, which is operated by AURA,
Inc., under NASA contract NAS5-26555.


\expandafter\ifx\csname natexlab\endcsname\relax\def\natexlab#1{#1}\fi

\newpage
\input{tab1}
\newpage
\input{tab2}

\newpage
\input{tab3}
\newpage
\input{tab4}

\newpage
\input{tab5}
\newpage
\input{tab6}
\newpage
\input{tab7}
\newpage

\newpage
%
%
\newpage
%
\newpage
%
\newpage

\end{document}

%% file: tab1.tex
\begin{table*}[t]
\caption[]{\label{t-log}{Log of observations.}}
\setlength{\tabcolsep}{0.3ex}
\begin{minipage}[t]{12.0cm}
\begin{flushleft}
\begin{tabular}[t]{lccccccccc}
\hline\noalign{\smallskip}
\hline\noalign{\smallskip}
System  & Date & & UT & Exp.\,[s] & State & $\rm F28\times50L$ & 
$V_{\rm range}$ & $P_{\rm
  orb}$\,[min]& B   \\
\hline\noalign{\smallskip}
V347 Pav           &   2003-03-28 &  & 02:23:05  &  700  &  low  & 17.7&15.2-17.6  & 90.0   & $<25$ \\
VV Pup             &   2003-03-10 &  & 10:17:34  &  900  &  low  & 17.4&14.5-18.0  & 100.4  & 31/56  \\
V834 Cen           &   2003-02-16 &  & 23:11:33  &  830  &  low  & 15.9&14.2-16.9  & 101.5  & 23    \\
CD Ind             &   2003-03-25 &  & 10:00:04  &  900  &  high & 15.8&16.2-17.4  & 110.9  & 11 \\
BL Hyi             &   2003-06-25 &  & 08:29:29  &  830  &  low  & 17.2&14.3-17.4  & 113.6  & 33  \\
MR Ser             &   2003-05-28 &  & 03:46:30  &  800  &  low  & 16.3&14.9-17.0  & 113.5  & 24    \\
ST LMi             &   2003-05-28 &  & 18:15:30  &  900  &  low  & 16.9&15.0-17.2  & 113.9  & 18\\
AN UMa             &   2003-06-23 &  & 13:09:16  &  900  &  high & 16.7&14.0-18.5  & 114.8  & 36  \\
UW Pic             &   2003-02-22 &  & 23:08:29  &  900  &  high & 15.8&16.4-17.2  & 133.4  & 19     \\
RX J1554           &   2003-02-27 &  & 23:18:13  &  830  &  low  & 16.7&15.0-17.0  & 151.9  & 145    \\
V895 Cen           &   2003-01-21 &  & 10:02:58  &  700  &  low  & 16.4&16.5-17.5  & 285.9  & ${\dag}$   \\
\hline\noalign{\smallskip}
\footnotesize{
$^{\dag}$ Not available}
\end{tabular}
\end{flushleft}
\end{minipage}
\end{table*}

%% file: tab2.tex
\centering
\begin{table*}[t]
\caption[]{\label{t-lines1} Line properties of the high-state magnetic CVs.}
\setlength{\tabcolsep}{1.2ex}
\begin{minipage}[t]{10.0cm}
\begin{flushleft}
\begin{tabular}[t]{l|cc|cc|cc}
\hline\noalign{\smallskip}
\hline\noalign{\smallskip}
\multicolumn{1}{l|}{} & \multicolumn{2}{|c|}{CD\,Ind} & \multicolumn{2}{|c|}{AN\,UMa} & \multicolumn{2}{|c}{UW\,Pic}\\
Lines & {\footnotesize EW} & {\footnotesize FWHM} & {\footnotesize EW} & {\footnotesize FWHM} & {\footnotesize EW} & {\footnotesize FWHM} \\
& $\rm \AA$ & $\rm \AA$ & $\rm \AA$ & $\rm \AA$ & $\rm \AA$ & $\rm \AA$ \\
%
\hline\noalign{\smallskip}
%
\HEiiL &  29 & 7 &  65 & 5 &  40 & 6 \\
\CivLL &  81 & 8 &  178 & 7 &  118 & 7 \\
\SIivLa &  17 & 6 &  24 & 6 &  14 & 5 \\
\SIivLb &  18 & 7 &  13 & 4 &  16 & 7 \\
\CiiLxiii &  7 & 6 &  15 & 7 &  4 & 6 \\
\SIiiiLL &  15 & 12 &  ...& ... &  ... &  ... \\
\NvLL &  22 & 9 &  42 & 8 &  13 & 8 \\ 
\SIiiiLa &  8 & 7 &  11 & 4 &  6 & 4 \\ 
\CiiiLxi &  17 & 6 &  10 & 3 &  9 & 6 \\
%
\hline\noalign{\smallskip}
\end{tabular}
\end{flushleft}
\end{minipage}
\end{table*}

%% file: tab3.tex
\centering
\begin{table*}[t]
\caption[]{\label{t-ratios} Line flux ratios of the high-state magnetic CVs.}
\setlength{\tabcolsep}{1.2ex}
\begin{minipage}[t]{10.0cm}
\begin{flushleft}
\begin{tabular}[t]{l|ccc}
\hline\noalign{\smallskip}
\hline\noalign{\smallskip}

Log line ratios & CD\,Ind & AN\,UMa & UW\,Pic\\
\hline\noalign{\smallskip}
\HEiil/\Civl &  -0.5 & -0.6 & -0.5  \\
\SIivl/\Civl & -0.3 & -0.6 & -0.6  \\
\Nvl/\Civl   & -0.4 & -0.7 & -0.8 \\
\hline\noalign{\smallskip}
\end{tabular}
\end{flushleft}
\end{minipage}
\end{table*}

%% file: tab4.tex
\begin{center}
\begin{table*}[t]
\caption[]{\label{t-lines2} 
Line properties of the low-state magnetic CVs.}
\setlength{\tabcolsep}{1.2ex}
\begin{minipage}[t]{10.0cm}
\begin{flushleft}
\begin{tabular}[t]{l|cc|cc|cc}
\hline\noalign{\smallskip}
\hline\noalign{\smallskip}
\multicolumn{1}{l|}{} & \multicolumn{2}{|c|}{V347\,Pav} & \multicolumn{2}{|c|}{VV\,Pup} & \multicolumn{2}{|c}{ST\,LMi} \\

Lines &  {\footnotesize EW} & {\footnotesize FWHM} & {\footnotesize
EW} & {\footnotesize FWHM} & {\footnotesize EW} & {\footnotesize FWHM}\\ 
& $\rm \AA$ & $\rm \AA$ & $\rm \AA$ & $\rm \AA$ & $\rm \AA$ & $\rm \AA$ \\
\hline\noalign{\smallskip}
%
\HEiiL    &  16 & 6  & ... & ...& 13  & 4   \\
\CivL     &  64 & 7  &   9 &  6 & 60  & 5   \\ 
\SIivLa   &  14 & 5  & ... & ...&  8  & 4   \\ 
\SIivLb   &   5 & 3  & ... & ...&  7  & 4   \\ 
\CiiLxiii &  19 & 8  & ... & ...& 28  & 6   \\ 
\SIiiiLa  &  34 & 7  & ... & ...& ... & ... \\
\CiiiLxi  &  24 & 4  & ... & ...& ... & ... \\
\hline\noalign{\smallskip}
\end{tabular}
\end{flushleft}
\end{minipage}
\end{table*}
\end{center}

%% file: tab5.tex
\begin{center}
\begin{table*}[t]
\caption[]{\label{t-models} Parameters obtained from model fits to
  the low-state spectra.}
\setlength{\tabcolsep}{1ex}
\begin{minipage}[c]{13.7cm}
\begin{flushleft}
\begin{tabular}[t]{l|ccc|cccc}
\noalign{\smallskip}
\hline\noalign{\smallskip}
\hline\noalign{\smallskip}
\multicolumn{1}{c|}{} & \multicolumn{3}{c|}{1-WD model}  & \multicolumn{4}{|c}{2-WD model$^{\footnotesize \rm c}$} \\
System & $N_1^{\footnotesize \rm a}$ & $T_1^{\footnotesize \rm a}$ &
$\chi_{\nu}^{2^{\footnotesize \rm b}}$ & $N_1+N_2^{\footnotesize \rm a}$ & $T_1^{\footnotesize \rm
  a}$ & $F^{\footnotesize \rm b}$ &  $\chi_{\nu}^{2^{\footnotesize \rm b}}$ \\
& [$10^{-23}\,\rm sr$] & [1000\,K] & & [$10^{-23}\,\rm sr$] & [1000\,K] & [$\%$] \\

\hline\noalign{\smallskip}
VV Pup   &  $2.85\pm{1.15\atop0.74}$ & $12.3\pm{0.7\atop0.7}$ & 1.6 & $4.33\pm{1.01\atop0.72}$  & $11.9\pm{0.6\atop0.5}$  & 0.02 &  1.5 \\ 
V834 Cen &  $3.47\pm{1.50\atop0.97}$ & $15.5\pm{1.0\atop1.0}$ & 4.1 & $4.78\pm{1.89\atop1.35}$  & $14.3\pm{0.9\atop0.9}$  & 0.31 &  1.4 \\ 
BL Hyi   &  $2.88\pm{1.43\atop0.84}$ & $14.2\pm{0.9\atop1.0}$ & 2.3 & $3.72\pm{1.59\atop1.07}$  & $13.3\pm{0.9\atop0.8}$  & 0.20 &  1.3 \\ 
MR Ser   &  $3.05\pm{1.18\atop0.85}$ & $15.1\pm{1.0\atop0.9}$ & 2.7 & $3.89\pm{1.58\atop1.12}$  & $14.2\pm{0.9\atop0.9}$  & 0.25 &  1.1 \\ 
V895 Cen &  $0.28\pm{0.11\atop0.08}$ & $15.3\pm{1.0\atop1.0}$ & 1.4 & $0.38\pm{0.15\atop0.11}$  & $14.0\pm{0.9\atop0.9}$  & 0.36 &  1.2 \\ 
V347 Pav &  $1.28\pm{0.53\atop0.38}$ & $13.3\pm{0.9\atop0.8}$ & 1.5 & $3.41\pm{0.77\atop0.61}$  & $11.8\pm{0.6\atop0.5}$  & 0.09 &  1.3 \\ 
ST LMi   &  $7.20\pm{1.86\atop1.52}$ & $10.9\pm{0.5\atop0.4}$ & 2.3 & $7.57\pm{1.82\atop1.79}$  & $10.8\pm{0.5\atop0.4}$  & 0.02 &  2.1 \\ 
\hline\noalign{\smallskip}
\end{tabular}
\end{flushleft}
\footnotesize 
$^{\rm a}$ $N_1$,and $N_2$, are the
 scale factors of the two-component fit (cold and hot respectively),
 and $T_1$ is the cold component
 temperature (or $\Twd$). The errors are due to the uncertainty in the
 mass \\
$^{\rm b}$ The $\chi^2_{\nu}$ and $F$ (fractional scale factor of the
 hot temperature component compared to the cold one) values are given
 only for the $\log g = 8.0$ case. Due to the nature of the
 observations (ACCUM mode), we are unable to properly quantify $F$,
 and therefore the values given here represents lower limits to the
 \textit{true} hot spot contribution\\
$^{\rm c}$ The temperature of the second component ($T_2$) is fixed at
50\,000\,K
\end{minipage}
\end{table*}
\end{center}

%% file: tab6.tex
\begin{center}
\begin{table*}[t]
\caption[]{\label{t-distances} Distances estimates to the low-state
magnetic CVs.}
\setlength{\tabcolsep}{0.5cm}
\begin{minipage}[t]{12.0cm}
\begin{flushleft}
\begin{tabular}[t]{l|c|c|c}
\hline\noalign{\smallskip}
\hline\noalign{\smallskip}
& 1-WD model & 2-WD model & Literature \\
System & $\rm d^{\footnotesize \dag}$\,[pc] & $\rm d^{\footnotesize \dag}$\,[pc] & d\,[pc] \\
\hline\noalign{\smallskip}
VV Pup   & $187\pm{22\atop33}$  & $151\pm{28\atop34}$ & $144^{\footnotesize \rm a}$\\
V834 Cen & $169\pm{18\atop28}$  & $144\pm{18\atop23}$ & $86^{\footnotesize \rm b}$ \\
BL Hyi   & $186\pm{15\atop30}$  & $163\pm{18\atop26}$ & $\sim130^{\footnotesize \rm c}$\\
MR Ser   & $180\pm{23\atop30}$  & $160\pm{18\atop26}$ & $139\pm13^{\footnotesize \rm d}$\\
V895 Cen & $600\pm{75\atop98}$  & $511\pm{60\atop81}$ & $250-295^{\footnotesize \rm  e}$\\
V347 Pav & $279\pm{31\atop43}$  & $171\pm{33\atop38}$ & $40-50^{\footnotesize \rm f}$\\
ST LMi   & $117\pm{22\atop23}$  & $115\pm{21\atop22}$ & $136^{\footnotesize \rm g}$ \\
\hline\noalign{\smallskip}
\end{tabular}
\end{flushleft}
\footnotesize 
$^\dag$ The errors are due to uncertainty in the mass \\
$^{\rm a}$\citet{bailey81-1}, $^{\rm b}$\citet{cropper90-1},
$^{\rm c}$\citet{beuermannetal85-1}, $^{\rm d}$\citet{schwopeetal93-1},
$^{\rm e}$\citet{howelletal97-2}, $^{\rm f}$\citet{ramsayetal04-1}, $^{\rm g}$\citet{baileyetal85-1}

\end{minipage}
\end{table*}
\end{center}

%% file: tab7.tex
\begin{center}
\begin{table*}[t]
\caption[]{\label{t-temperatures} List of CV\,WD temperatures}
\setlength{\tabcolsep}{4.8ex}
\begin{minipage}[c]{17cm}
\begin{flushleft}
\begin{tabular}[t]{lccc}
\noalign{\smallskip}
\hline\noalign{\smallskip}
\hline\noalign{\smallskip}
System & $P_{\rm orb}$[min] & $T_{\rm wd}$[K] & Ref.\\
\hline\noalign{\smallskip}
BW\,Scl             &      78.2       &      14\,800  & 1   \\
LL\,And             &      79.2       &      14\,300  & 2   \\
EF\,Eri$^{\ddag}$   &      81.0       &      9\,500   & 3   \\
WZ\,Sge             &      81.6       &      14\,900  & 4   \\
AL\,Com             &      81.6       &      16\,300  & 5   \\
SW\,UMa             &      81.8       &      14\,000  & 1   \\
HV\,Vir             &      83.5       &      13\,300  & 6   \\
WX\,Cet             &      83.9       &      14\,500  & 7   \\
EG\,Cnc             &      86.4       &      13\,300  & 6   \\
DP\,Leo$^{\ddag}$   &      89.8       &      13\,500  & 8  \\
V347\,Pav           &      90.0       &      12\,300  & $\dag$ \\
BC\,UMa             &      90.2       &      15\,200  & 1   \\
VY\,Aqr             &      90.8       &      14\,000  & 7  \\
EK\,Tra             &      91.6       &      18\,800  & 9  \\
VV\,Pup             &      100.4      &      12\,100  & $\dag$ \\
V834\,Cen           &      101.5      &      14\,200  & $\dag$ \\
VW\,Hyi             &      106.9      &      19\,000  & 10  \\
CU\,Vel             &      113.0      &      18\,500  & 11  \\
BL\,Hyi             &      113.6      &      13\,100  & $\dag$ \\
MR\,Ser             &      113.5      &      14\,000  & $\dag$ \\
ST\,LMi             &      113.9      &      10\,800  & $\dag$ \\
EF\,Peg             &      123        &      16\,600  & 2  \\
HU\,Aqr$^{\ddag}$   &      125.0      &      14\,000  & 12  \\
QS\,Tel$^{\ddag}$   &      139.9      &      17\,500  & 13  \\
AM\,Her$^{\ddag}$   &      185.6      &      20\,000  & 14  \\
TT\,Ari             &      198.0      &      39\,000  & 15     \\
DW\,UMa             &      198.0      &      50\,000  & 16     \\
MV\,Lyr             &      191.0      &      47\,000  & 17    \\
V1043\,Cen$^{\ddag}$&      251.4      &      15\,000  & 18  \\
U\,Gem              &      254.7      &      31\,000  & 19,20 \\
SS\, Aur            &      263.2      &      31\,000  & 21  \\
V895\,Cen           &      285.9      &      13\,800  & $\dag$ \\
RX\, And            &      302.2      &      34\,000  & 22  \\
Z\,Cam              &      417.4      &      57\,000  & 23 \\
RU\,Peg             &      539.4      &      49\,000  & 21  \\
\hline\noalign{\smallskip} 
\end{tabular}
\end{flushleft}
\footnotesize
$^1$\citet{gaensickeetal04-2}, 
$^2$\citet{howelletal02-1},
$^3$\citet{beuermannetal00-1}, 
$^4$\citet{sionetal95-2},
$^5$\citet{szkodyetal03-1} 
$^6$\citet{szkodyetal02-1}, 
$^7$\citet{sionetal03-1},
$^8$\citet{schwopeetal02-1},
$^9$\citet{gaensickeetal01-3},
$^{10}$\citet{gaensicke+beuermann96-1},
$^{11}$\citet{gaensicke+koester99-1},
$^{12}$\citet{gaensicke99-1},
$^{13}$\citet{rosenetal01-1},
$^{14}$\citet{gaensickeetal95-1},
$^{15}$\citet{gaensickeetal99-1},
$^{16}$\citet{araujo-betancoretal03-1},
$^{17}$\citet{hoardetal04-1},
$^{18}$\citet{gaensickeetal00-1},
$^{19}$\citet{sionetal98-1},
$^{20}$\citet{long+gilliland99-1},
$^{21}$\citet{sionetal04-1},
$^{22}$\citet{sionetal01-1},
$^{23}$\citet{hartleyetal04-1},
$^{\ddag}$Polars whose temperatures were found in the literature,
$^{\dag}$Temperatures obtained in this work (see Table\,\ref{t-models})
\end{minipage}
\end{table*}
\end{center}